\documentclass[superscriptaddress,amsmath,amssymb, aps, prl,longbibliography,nofootinbib,twocolumn]{revtex4-2}
\usepackage[normalem]{ulem}
\usepackage{float}
\usepackage{bm}
\usepackage{xcolor}
\usepackage[dvipsnames]{xcolor}
\usepackage{mathtools}  
\usepackage{graphicx}
\usepackage{hyperref}

\hypersetup{colorlinks,breaklinks,
            urlcolor=[rgb]{0.08,0.03,1},
            linkcolor=[rgb]{0.08,0.03,1},
            citecolor=[rgb]{0.08,0.03,1}}

\usepackage{booktabs} % better tables

\newcommand{\tfr}{t}
\newcommand{\bfr}{b}
\newcommand{\kk}{\mathbf{k}}
\newcommand{\stiff}{\rho_s}

\makeatletter
\renewcommand\footnotesize{%
   \@setfontsize\footnotesize\@ixpt{11}%
   \abovedisplayskip 1\p@ \@plus2\p@ \@minus4\p@
   \abovedisplayshortskip \z@ \@plus\p@
   \belowdisplayshortskip 2\p@ \@plus2\p@ \@minus2\p@
   \def\@listi{\leftmargin\leftmargini
               \topsep 2\p@ \@plus2\p@ \@minus2\p@
               \parsep 1\p@ \@plus\p@ \@minus\p@
               \itemsep \parsep}%
   \belowdisplayskip \abovedisplayskip
}
\makeatother

\newcommand{\Harvard}{Department of Physics, Harvard University, Cambridge, Massachusetts 02138, USA.}
\newcommand{\TUM}{Technical University of Munich, TUM School of Natural Sciences, Physics Department, 85748 Garching, Germany}
\newcommand{\MCQST}{Munich Center for Quantum Science and Technology (MCQST), Schellingstr. 4, 80799 M{\"u}nchen, Germany}

\begin{document}

\title{False Vacuum Decay in Flat-Band Ferromagnets: Role of Quantum Geometry and Chiral Edge States}

 \author{Fabian Pichler}
 \thanks{These authors contributed equally to this work.\\}
 \affiliation{\TUM}
 \affiliation{\MCQST}
 \author{Clemens Kuhlenkamp}
 \thanks{These authors contributed equally to this work.\\}
 \affiliation{\Harvard}
 \author{Michael Knap}
 \affiliation{\TUM}
 \affiliation{\MCQST}

\date{\today}

\begin{abstract}

Dynamical control of quantum matter is a challenging, yet promising direction for probing strongly correlated states. Motivated by recent experiments in twisted MoTe$_2$ that demonstrated optical control of magnetization, we propose a protocol for probing magnetization dynamics in flat-band ferromagnets. We investigate the nucleation and dynamical growth of magnetic bubbles prepared on top of a \textit{false vaccum} in both itinerant ferromagnets and spin-polarized Chern insulators. For ferromagnetic metals, we emphasize the crucial role of a non-trivial quantum geometry in the magnetization dynamics, which in turn also provides a probe for the quantum metric. Furthermore, for quantum Hall ferromagnets, we show how properties of chiral edge modes localized at domain-wall boundaries can be dynamically accessed. Our work demonstrates the potential for nonequilibrium protocols to control and probe strongly correlated phases, with particular relevance for twisted MoTe$_2$ and graphene-based flat-band ferromagnets. 
\end{abstract}

\maketitle

\begin{figure*}
\begin{center}
\includegraphics[width=0.99\linewidth]{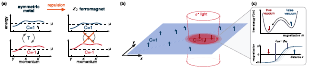}
\caption{\textbf{Bubbles, domain walls, and false vacuum decay.} \textbf{(a)} Itinerant Ising-magnetism in two-dimensional topological bands. Strong correlations drive the $\mathbb{Z}_2$ symmetric metal (left) into a spin-polarized ferromagnetic phase (right), which allows the system to minimize its exchange energy via the Stoner mechanism. Depending on the filling, both quantum Hall ferromagnets and itinerant magnets can be stabilized. \textbf{(b)} In materials such as twisted-MoTe$_2$ bilayers, the magnetization can be optically controlled by impinging the sample with circularly polarized light. \textbf{(c)} An external bias magnetic field $B_z$ prepares the system  in a  metastable \emph{false vacuum}. We find that the growth dynamics of a bubble in the \emph{true vacuum}, that can be characterized optically by measuring magnetization outside the original bubble, depend strongly on the underlying electronic state. }
\label{fig:overview}
\end{center}
\end{figure*}

\textbf{{Introduction.---}}Spontaneous time-reversal symmetry breaking and the emergence of ferromagnetism driven by strong electronic correlations dates back to early work by Stoner~\cite{stonerCollectiveElectronFerromagnetism1938, Shimizu_1981}. Yet, two-dimensional Stoner ferromagnetism has only recently been experimentally observed in van der Waals heterostructures~\cite{Wu_TopologicalInsulators_2019,sharpeEmergentFerromagnetismThreequarters2019, liQuantumAnomalousHall2021a, Zhou_IsospinMagnetism_2022, 
hanCorrelatedInsulatorChern2024, anderson_ProgrammingCorrelated_2023, cai_SignaturesFractional_2023, zeng_ThermodynamicEvidence_2023, park_ObservationFractionally_2023, xu_ObservationInteger_2023, LongJu2024, xieTunableFractional2025, gao2025probing}. In these systems, approximately flat electronic bands enhance electronic correlations, favoring spontaneous spin polarization and stabilizing various correlated phases, such as quantum anomalous Hall states~\cite{Ghaemi2012, Bultinck2020, Repellin2020Ferro, Wu2020_QuantumGeometry, Li2021Spontaneous, devakulMagicTwistedTransition2021, Crepel2023FlatBandFerro, kang2024quantumgeometricboundsaturated, chen2025quantumgeometricdipoletopologicalboost}. This experimental progress has also renewed theoretical interest in the nature of Stoner transitions and itinerant ferromagnetism in two dimensions~\cite{Chubukov2008, Zhu2018ValleyStoner, Valenti2024Nematic, Raines2024Unconventional, Raines2024Stoner, Mayrhofer2025Stoner, Calvera2025}.
Despite these advances, understanding various static and dynamical aspects of Stoner ferromagnetism remains largely an open question.

In this work, we propose 2D quantum materials as a promising platform for exploring nonequilibrium dynamics in the presence of strong underlying electronic correlations. 
Building on recent experiments in twisted MoTe$_2$~\cite{huberOpticalControlTopological2025, holtzmannOpticalControlInteger2025, caiOpticalSwitchingMoire2025}, we establish a protocol for observing magnetization dynamics in flat-band ferromagnets. 
Concretely, we consider an Ising magnet in a given polarization that is energetically slightly disfavored by a weak magnetic field, thereby realizing a metastable \textit{false vacuum}. Creating magnetic bubbles in the \textit{true vacuum}, initializes a nonequilibrium magnetization dynamics that can be interpreted as \textit{false vacuum decay}~\cite{Coleman_FalseVacuum_1977}, that we show to sensitively depend on the underlying electronic state. Our theory applies to the dynamics of magnetic domains in itinerant magnets as well as that of gapped quantum Hall magnets. 
This not only enables studying nucleation dynamics in quantum materials~\cite{Lagnese2021, Milsted2022, shavitEphemeralSuperconductivityAtop2025}, but also highlights how signatures of the underlying correlated states can be extracted from such dynamics~\cite{Qiu2025_excitons}.

\textbf{{Phenomenology of bubble dynamics.---}}Motivated by the optical control over ferromagnetism in twisted MoTe$_2$ established in~\cite{huberOpticalControlTopological2025, holtzmannOpticalControlInteger2025, caiOpticalSwitchingMoire2025}, we propose the following protocol to investigate real-time dynamics in a two-dimensional Ising-ferromagnet (see Fig.~\ref{fig:overview}a,b): First, apply a weak magnetic field $B_z$ to fix a polarization of the sample. Since we are working within the ferromagnetic phase, the polarization persists even after the magnetic field is turned off. Next, return to zero magnetic field and optically flip the magnetization in a finite region of the sample, thus creating a bubble of radius $R$ with opposite magnetization and Berry curvature. When no external field is applied, both polarizations are degenerate in energy, and domains will undergo coarsening dynamics. However, this degeneracy can be explicitly broken by applying an external magnetic field $B_z$ in the opposite direction to the initial polarization. Let us consider the difference in free-energy density between the two polarizations
$\Delta f\sim B_z$, so that the $\uparrow$-polarization is metastable and in a so-called false-vacuum state (see Fig.~\ref{fig:overview}c). 
Whether or not the magnetic bubble in the true vacuum will dynamically grow is then determined by a competition between a gain in bulk energy $\Delta f$ at the cost of an increase in surface energy, determined by the surface tension $\sigma$ between magnetic domain walls. 

From this energy balance, we deduce a typical critical radius of
\begin{equation}
    R_c = \frac{\sigma}{\Delta f} \label{eq:citicalRadius}
\end{equation}
above which the bubble will grow, and below which it will shrink. By performing dynamical experiments, the domain-wall surface tension is obtained from determining $R_c$. The critical radius $R_c$ in turn can be extracted from the early-time magnetization dynamics by changing the optical spot size at fixed $B_z$, or by keeping the spot size fixed and tuning $\Delta f$  magnetically. Crucially, the surface tension is an intricate function of the underlying system parameters and unveils properties of the quantum geometry and chiral edge states, as we will show below.

\begin{figure}
\begin{center}
\includegraphics[width=0.75\linewidth]{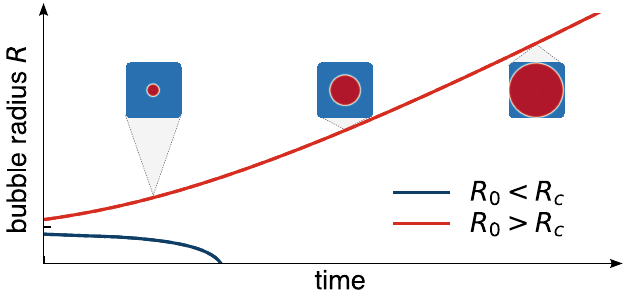}
\caption{\textbf{Dynamics of a nucleated bubble.} The size of the initial domain $R_0$ relative to the critical bubble radius $R_c$, determines the fate of the metastable system. If $R_0<R_c$ (blue line), the surface tension will dominate and the bubble will shrink, leaving the meta-stable \textit{false vacuum} intact. By contrast, for $R_0>R_c$ (red line) the bubble will expand asymptotically with a constant velocity $v_B$ driving the system to the lowest energy configuration.
}
\label{fig:bubble}
\end{center}
\end{figure}

We consider a phenomenological treatment of generic Ising magnets, whose Landau-Ginzburg free energy is given by
\begin{equation}
\begin{aligned}
    f[m] \approx f_0 +  \frac{\stiff}{2}(\nabla m)^2 + \frac{r_0}{2} m^2  + \frac{\lambda}{4!}m^4,
    \label{eq:LG_free_energy}
\end{aligned}
\end{equation}
where $m = n_\uparrow-n_\downarrow$ is the magnetization, $\stiff$ the spin stiffness, and $r_0$ and $\lambda$ are the mass and self-interaction terms, respectively. 
In the long-time limit, the order-parameter dynamics are damped and relax with an equilibration rate $\Gamma$. Experimentally, this rate arises from intrinsic many-body interactions, disorder, as well as from external baths, such as phonons or the coupling to leads. These processes do not conserve the total magnetization $n_\uparrow-n_\downarrow$, and at late times the order parameter follows `model-A' dynamics~\cite{HohenbergHalperin77}:
\begin{equation}
    \partial_t m(\mathbf{x},t) = -\Gamma\frac{\delta f[m]}{\delta m(\mathbf{x},t)} + \xi(\mathbf{x},t),
    \label{eq:model_a}
\end{equation}
where $\xi$ encodes fluctuations of a bath whose strength depends on temperature.
The dynamics of a nucleated domain follow from Eq.~\eqref{eq:model_a}, and generic initial magnetization profiles relax quickly to equilibrium domain-wall configurations, similar in size to the initialized domain. Assuming that the domain-wall width $\ell_\text{DW}$ is much smaller than its size and that the applied magnetic fields are weak, the dynamics of a bubble with radius $R(t)$ is
\begin{equation}
    \dot{R}(t) = v_B - \frac{\stiff\Gamma}{R(t)}, 
    \quad v_B=\Delta f \;\frac{\Gamma \stiff}{\sigma},
    \label{eq:bubblegrowth}
\end{equation}
where $\Delta f\simeq g_h\mu_B B_z m$.
Thus, at late times supercritical bubbles grow with constant velocity $v_B$ (see Fig.~\ref{fig:bubble})~\cite{Allen1979,Chaikin_Lubensky_1995}.
The empirically observed relaxation rate $\Gamma$ can be slow in semiconductor heterostructures, particularly at low temperatures~\cite{Smolenski_relax_2022}, which enables the experimental observation of the nucleation dynamics. When determining the relaxation rate from an independent relaxation measurement, $v_B$ directly encodes information about the surface tension $\sigma$ and spin stiffness $\stiff$.
Ideally, measurements of both $R_c$ and $v_B$ are combined, to provide robust estimates on $\sigma$ and $\stiff$.

We apply the theory for bubble dynamics first in the case of itinerant narrow-band ferromagnets, where we show that the Ginzburg-Landau action describes itinerant Stoner magnets. We then analyze gapped quantum Hall ferromagnets, where we show that chiral edge modes give relevant contributions to the surface tension.

{\textbf{Surface tension of itinerant flat-band ferromagnets.---}}The bubble dynamics governed by Eq.~\eqref{eq:model_a} sensitively depend on microscopic details of the underlying quantum state. To analyze domain walls in itinerant magnets, we consider the Ginzburg-Landau free energy of Eq.~\eqref{eq:LG_free_energy}. 
The magnetization profile is then given by
\begin{equation}
    m(z) = m_0 \tanh\Big(\frac{z}{\ell_\mathrm{DW}}\Big), \quad \text{with} \quad \ell_\mathrm{DW} = \sqrt{\frac{2\stiff}{-r_0}}
\end{equation}
and $m_0 = \sqrt{-6r_0/\lambda}$. The surface tension associated with the domain wall can be estimated to be
\begin{equation}
    \sigma = \frac{2}{3} (-r_0) m_0^2 \ell_\mathrm{DW}. \label{eq:surfacetension}
\end{equation}

Starting out with a microscopic electronic model of an itinerant Ising-ferromagnet, we estimate the free energy by expanding the action in powers of the magnetization $m$ at finite temperature, yielding to lowest order
\begin{equation}
\begin{aligned}
    r_0 &= 2U\big[1-U \Pi(\mathbf{q=0})\big] ,\\
    \stiff &= 2U^2 \big[\partial_\mathbf{q}^2 \Pi(\mathbf{q})\big]_{\mathbf{q}=0} \label{eq:LG_parameters}
\end{aligned}
\end{equation}
where $\Pi(\mathbf{q})$ is the static particle-hole bubble and $U$ sets the interaction strength. 
For quadratically dispersing electrons $E(\mathbf{k}) = \mathbf{k}^2/2m_e^*$ in two dimensions, Eq.~\eqref{eq:LG_parameters} suggests $\stiff=0$ and therefore an inevitable breakdown of a description in terms of Eq.~\eqref{eq:LG_free_energy}. In bands with non-trivial quantum geometry, however, we show that the particle-hole bubble acquires a finite zero-momentum curvature~\cite{Chen2023, Iskin2023, Kitamura2024, Shavit_2025, supp}
\begin{equation}
    \Pi(\mathbf{q}) = \Pi_0 \big[1-\bar{g}^\mathrm{FS}_{ab} q_a q_b + \mathcal{O}(q^3)\big], \label{eq:staticPolarization}
\end{equation}
such that the stiffness $\stiff$, and therefore also the surface tension $\sigma$, is generically proportional to the Fermi-surface averaged quantum metric $\bar{g}_\mathrm{FS} = \mathrm{Tr}[\bar{g}_{ab}^\mathrm{FS}]$. Furthermore, as highlighted by Refs.~\cite{Wu2020_QuantumGeometry, kang2024quantumgeometricboundsaturated, kitamura2025quantumgeometricferromagnetismsingular}, a finite quantum metric is essential for controlling fluctuations near the critical point.
Importantly, from Eqs.~(\ref{eq:LG_parameters},~\ref{eq:staticPolarization}) it follows that the domain-wall width $\ell_\mathrm{DW} \propto \sqrt{\bar{g}_\mathrm{FS}}$. Consequently, the surface tension also sensitively depends on the quantum metric, which is bound by the Berry curvature $\mathrm{Tr}[g_{ab}(\mathbf{k})] \geq |\mathcal{B}(\mathbf{k})|$, guaranteeing also non-trivial geometry in topologically non-trivial bands~\cite{roy2014}. This highlights the importance of a finite Berry curvature in flat-band ferromagnets, such as twisted MoTe$_2$~\cite{Wu2020_QuantumGeometry, kang2024quantumgeometricboundsaturated, kitamura2025quantumgeometricferromagnetismsingular}.

\begin{figure}
\begin{center}
\includegraphics[width=0.99\linewidth]{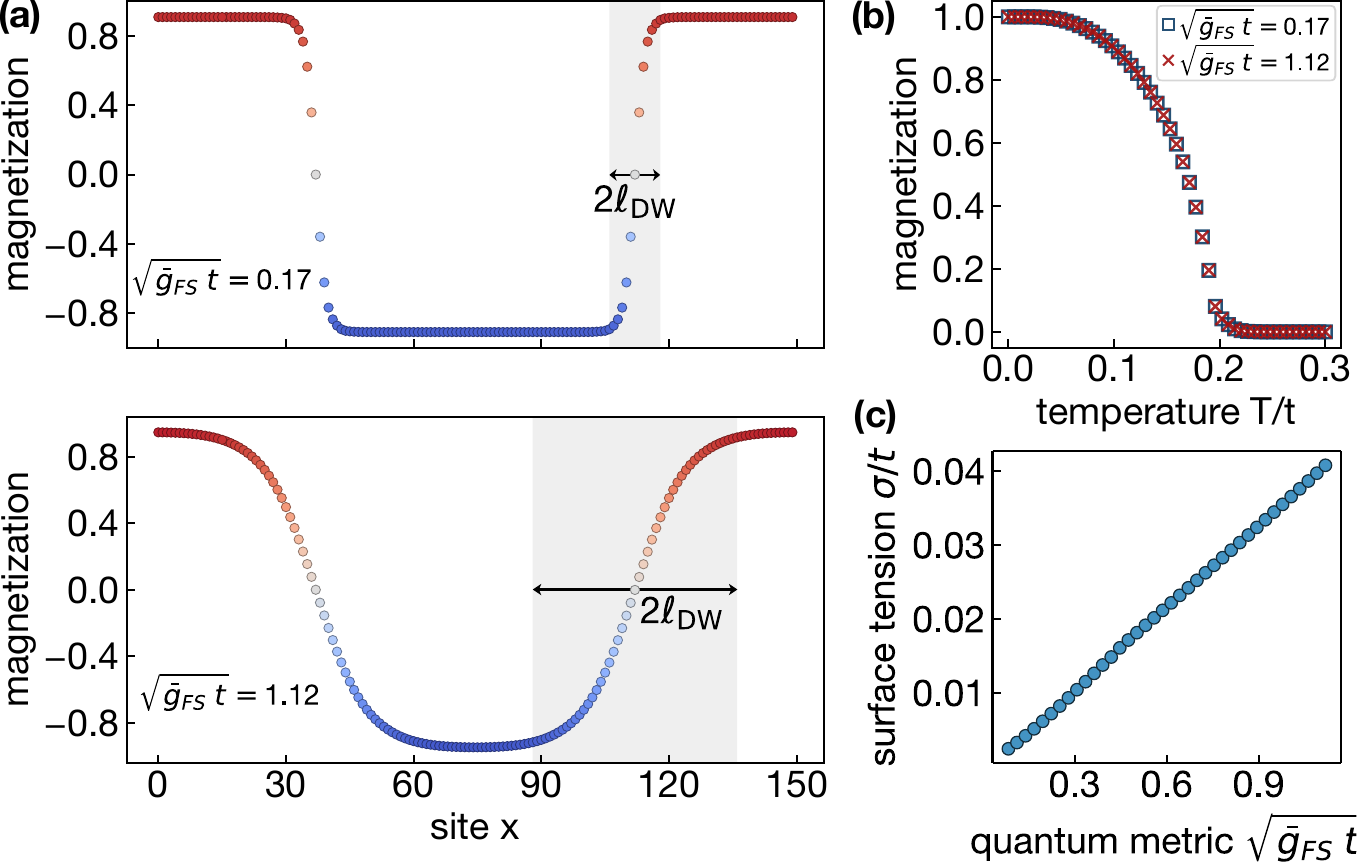}
\caption{\textbf{Consequence of quantum metric on the domain wall formation in itinerant ferromagnets.} \textbf{(a)} Magnetization across two domain walls for two different values of the quantum metric; corresponding to $\zeta=3$ and $\zeta=20$ in Model~\eqref{eq:TMmodel}. \textbf{(b)} Normalized magnetization as function of temperature. The critical temperature does not depend on the quantum metric, while the domain wall structure does. \textbf{(c)} Linear dependence of the surface tension on the square root of the Fermi-surface averaged quantum metric $\sqrt{\bar{g}_\mathrm{FS}}$, with a fixed quadratic dispersion. All lengths are given in units of the lattice constant $a$.}
\label{fig:itinerant}
\end{center}
\end{figure}

To numerically check our analytical results on quantum-geometry dependence of the surface tension, we consider the tunable metric model (TM model; see End Matter Eq.~\eqref{eq:TMmodel}) on a square lattice
whose quantum geometry can be tuned independently of its band dispersion~\cite{hofmannHeuristicBoundsSuperconductivity2022, hofmannTMmodel2023, Shavit_2025}. This conveniently allows us to study the influence of quantum geometry on domain walls, without the interference of band-structure effects. 
We are interested in the structure of the domain walls between two ferromagnetic domains, which we determine using self-consistent mean-field theory at finite temperatures. Technical details are presented in the End Matter. As predicted from the Stoner criterion, we find the critical temperature to only depend on the density of states at the Fermi energy and the interaction strength $U$ (Fig.~\ref{fig:itinerant}b). 
Conversely, the domain-wall structure and surface tension sensitively depend on the Fermi-surface averaged quantum metric $\bar{g}_\mathrm{FS}$ (Fig.~\ref{fig:itinerant}a,c). 
We find the surface tension to grow linearly with the square root of the quantum metric $\sigma \propto\sqrt{\bar{g}_\mathrm{FS}}$, as predicted by Eqs.~(\ref{eq:surfacetension},~\ref{eq:LG_parameters},~\ref{eq:staticPolarization}). The bubble growth velocity $v_B$ Eq.~\eqref{eq:bubblegrowth} and the critical radius $R_c$ Eq.~\eqref{eq:citicalRadius} inherit the same dependence on the quantum metric. 
We have cross-checked that our Landau-Ginzburg treatment is quantitatively consistent with self-consistent Hartree-Fock calculations; see supplemental material~\cite{supp}.

Although thermal fluctuations are important for Ising transitions in two dimensions, the temperature at which they start modifying the mean-field result can be estimated within the Landau-Ginzburg framework of Eq.~\eqref{eq:LG_free_energy} by the Ginzburg temperature $T_G$~\cite{Chaikin_Lubensky_1995}
\begin{equation}
    t_G = \frac{|T_G-T_c|} {T_c} \approx 0.4\frac{T_c}{\pi \stiff\, m_0^2}. \label{eq:relativeGinzburgTemp}
\end{equation}
When the ratio $T_c/(\stiff m_0^2)$ is small, $T_G$ is close to $T_c$, in which case mean-field behavior can be observed over a wide temperature regime. 
Crucially, a large quantum metric increases the spin stiffness $\stiff$, thus controlling quantum fluctuations and lowering the reduced Ginzburg temperature $t_G$. As a concrete example, we estimate the reduced Ginzburg temperature for twisted MoTe$_2$ and find it to be small $t_G \sim \mathcal{O}(10^{-2} - 10^{-4})$, justifying our mean-field treatment (see End Matter Fig.~\ref{fig:GinzburgTemperatureMoTe2}).

\begin{figure}
\begin{center}
\includegraphics[width=0.99\linewidth]{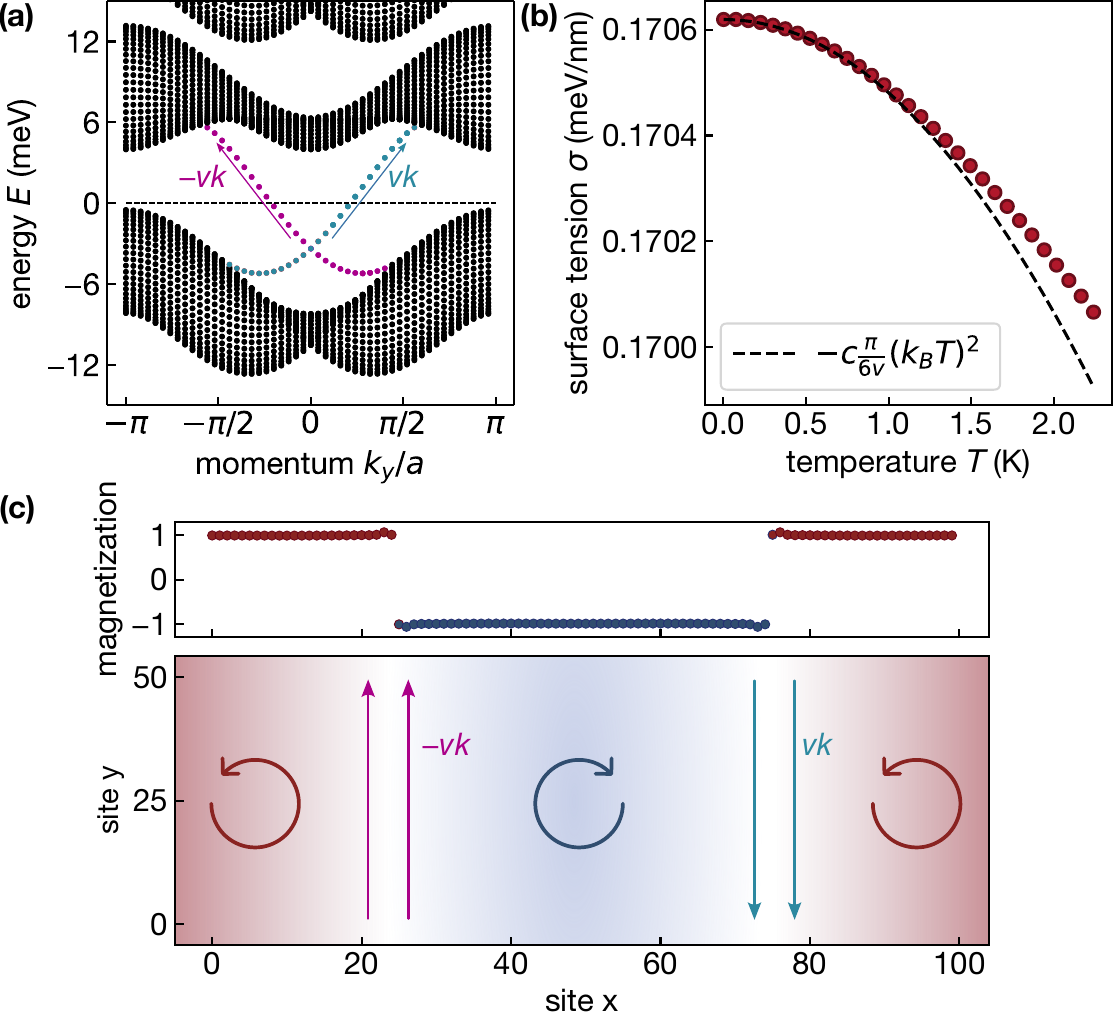}
\caption{\textbf{Edge mode contribution to the domain-wall surface tension.}  \textbf{(a)} Self-consistently determined spectrum of a $\nu=1$ Chern insulator with two domain walls, leading to four chiral edge modes (two per domain wall), with linear dispersion $\pm v k$ around the Fermi surface. \textbf{(b)} Temperature dependence of the surface tension, which for small $T$ is dominated by entropic contributions from the edge modes, Eq.~\eqref{eq:edgemodecontribution}. 
\textbf{(c)} Magnetization across the system (top) and sketch of domains (bottom). At the domain wall, where edge modes are present, the magnetization jumps discontinuously.
}
\label{fig:chiral}
\end{center}
\end{figure}

\textbf{{Chiral edge mode contribution to surface tension.---}}In topologically non-trivial bands, incompressible quantum anomalous Hall states can form at special fillings $\nu\in\{1, \; \frac{2}{3},\;\dots\}$ by spin (and valley) polarization~\cite{Ghaemi2012, Bultinck2020, Repellin2020Ferro, Wu2020_QuantumGeometry,Li2021Spontaneous, devakulMagicTwistedTransition2021, Crepel2023FlatBandFerro, kang2024quantumgeometricboundsaturated,chen2025quantumgeometricdipoletopologicalboost}. The transition between a spin-unpolarized chiral metal and a spin-polarized quantum Hall state is generically expected to be of first order, limiting a controlled Landau-Ginzburg approach. However, we can nonetheless derive universal predictions arising from the gapless chiral edge modes that are enforced between domain walls of quantum Hall states of opposite chirality (see Fig.~\ref{fig:chiral}a)~\cite{Qiu2025}. At temperatures $k_BT \ll \Delta$ below the quantum Hall gap $\Delta$, the free energy of the theory depends sensitively on the edge modes. Consistent with expectations from conformal field theory~\cite{Bloete1986}, we find that they contribute a term
\begin{equation}
    f(T)_{1+1d} = -c \;\frac{\pi}{6 v} \; (k_BT)^2, \label{eq:edgemodecontribution}
\end{equation}
where $c$ is the central charge of the gapless edge-modes localized at the domain wall and $v$ is their velocity, see also Ref.~\cite{Shavit_2022}. Since there are two chiral edge modes per domain wall, we have $c=1$.

To estimate the surface tension and edge-mode velocity, we consider a twisted MoTe$_2$ inspired honeycomb-lattice model and focus on integer filling $\nu=1$ (see End Matter for the model). We include up to third-nearest neighbor hopping with parameters adopted from Ref.~\cite{chen2025fractionalcherninsulatorquantum}, approximating the dispersion of twisted MoTe$_2$. For simplicity, we only include on-site interactions. 
This allows us to provide a qualitative estimate of the surface tension. 
For a magnetic field of $B_z = 10\; \mathrm{mT}$, we find the critical bubble radius to be $R_c \sim 2 \; \mu\mathrm{m}$, in the same range as typical optical spot sizes.
For small temperatures, the surface tension is accurately described by $\sigma(T) \approx \sigma_0 + f(T)_{1+1d}$ (see Fig.~\ref{fig:chiral}b). Hence, by measuring the temperature dependence of the surface tension, it is possible to directly extract the velocity of the chiral edge mode. 
Here, we focused on quantum Hall states with a single chiral edge mode, corresponding to two modes per domain wall~\cite{Qiu2025}. Systems with higher Chern number~\cite{LedwithPRL2022, Niu2025} or fractional quantum Hall states can host several edge modes, with possibly different velocities $v_i$ and central charge $c_i$. In that case, each mode contributes independently a term $f(T)_{1+1d}$ to the surface tension and thus, an averaged edge velocity can be extracted.

Compared to the smooth domain walls for itinerant ferromagnets (see Fig.~\ref{fig:itinerant}), we obtain very narrow, discontinuous domain walls for the $\nu=1$ quantum Hall state. This is a consequence of the gapped bulk. Due to the presence of chiral edge modes, the particle density (and consequently also the magnetization) is even slightly enhanced (Fig.~\ref{fig:chiral}c), highlighting the intricate microscopic structure of domain walls in quantum Hall states.

\textbf{{Experimental implications \& Outlook.---}}In this work, we have presented a simple experimental protocol to study nucleation dynamics and false vacuum decay in two-dimensional quantum ferromagnets and uncovered the role of quantum geometry and chiral edge modes on the domain wall formation. Our theory immediately applies to twisted MoTe$_2$, in which the optical control of magnetization has been demonstrated experimentally~\cite{huberOpticalControlTopological2025, holtzmannOpticalControlInteger2025, caiOpticalSwitchingMoire2025}.
To detect false vacuum decay, we propose to measure magnetization dynamics using low-power circularly polarized light outside the initially created bubble. 
When spin dynamics is sufficiently slow, time-resolved measurements of magnetization dynamics become feasible, particularly when combined with super-resolution microscopy. This enables studies of the effects of coupling to gapless fermions, edge reconstruction at fractional fillings, and the influence of disorder on domain wall dynamics, which are interesting directions for future work. 

Our results showed that the stiffness of two-dimensional itinerant Stoner magnets is controlled by quantum geometry. In specific materials, contributions due to non-quadratic band-structure effects will quantitatively alter the stiffness. For quantitative estimates detailed theoretical modeling of the relevant microscopic system will be pertinent. %Nevertheless, we expect the quantum metric to provide the leading-order contribution. 
Quantum metric effects will be especially striking in materials with highly localized Berry curvature, such as rhombohedral graphene~\cite{DongRhomboGraphene2024, DongJunkaiRhomboGraphene2024, DongPatri2024}, where our theory predicts a strong density dependence of the domain wall dynamics.

Tuning the quantum geometry opens up new experimental and theoretical pathways to explore and control Stoner transitions in two dimensions, whose properties are still open to a large extent. Provided the Stoner transition remains continuous all the way to zero temperature, quantum fluctuations of the magnetization will couple to itinerant electrons and strongly alter their transport properties. It will be exciting to investigate such effects in experiments, where both magnetization and transport are readily accessible.

Furthermore, materials such as twisted MoTe$_2$, that couple strongly to light, can be driven far from thermal equilibrium using optical techniques. In the presence of strong drives, we expect that magnetic order can be stabilized even when the system starts out in a trivial paramagnetic state, thus enabling the realization of light-induced topological states. 
Understanding the underlying time scales and the experimental implications of such light-induced states is another exciting future direction.

%\textit{\textbf{Note added.---}}While finalizing this manuscript, we became aware of a related theoretical work~\cite{breach2026}.

\textbf{{Acknowledgments.---}}We thank J. Dong, J. Feldmeier, A. Imamoglu, K. Klocke, P. J. Ledwith, C. Zelle, and T. Smolenski for insightful discussions. C.K. acknowledges funding from the Swiss National Science Foundation (Postdoc.Mobility Grant No. 217884). 
 F.P. and M.K. acknowledge support from the Deutsche Forschungsgemeinschaft (DFG, German Research Foundation) under Germany’s Excellence Strategy–EXC–2111–390814868, TRR 360 – 492547816 and DFG grants No. KN1254/1-2, KN1254/2-1, the European Research Council (ERC) under the European Union’s Horizon
2020 research and innovation programme (grant agreement No 851161), the European Union (grant agreement No 101169765), as well as the Munich Quantum Valley, which is supported by the Bavarian state government with funds from the Hightech Agenda Bayern Plus.  

\textbf{{Data availability.---}}Data, data analysis, and simulation codes are available upon reasonable request on Zenodo~\cite{zenodo}.

\addtocontents{toc}{\string\tocdepth@munge}

\bibliography{library}
\appendix

\begin{center}
    \textbf{End matter}
\end{center} 

\renewcommand{\thefigure}{A\arabic{figure}}
\setcounter{figure}{0}
\section{Domain-wall structure from self-consistent mean-field theory}\label{app:selfconsistentMF}

We perform self-consistent mean-field calculations to obtain the domain-wall structure~\cite{wang2022structuredomainwallschiral}. We consider a system with $L_x \times L_y$ unit cells and periodic boundary conditions in both directions.
\begin{equation}
    H = H_0 + U \displaystyle\sum_{\substack{x, y \\ a}} c^\dagger_{a\uparrow}(x, y) c^\dagger_{a\downarrow}(x, y) c_{a\downarrow}(x, y) c_{a\uparrow}(x, y) \label{eq:generalHamiltonian}
\end{equation}
where $H_0$ is the kinetic term (describing either the tunable metric model or an effective model for twisted MoTe$_{2}$) and $a$ is some sublattice or orbital degree of freedom. We mean-field decouple the interactions and assume translational invariance along the $y$ direction, allowing us to Fourier transform in that direction. This leads to the following mean-field Hamiltonian
\begin{equation}
    \begin{aligned}
        H_\mathrm{MF} = H_0 +  U& \displaystyle\sum_{k = 1}^{L_y} \displaystyle\sum_{x, a, \sigma} \big[ \chi_{aa, \sigma\sigma}(x) c^\dagger_{a \bar{\sigma} k}(x) c_{a \bar{\sigma} k}(x) \\ &\quad- \chi_{aa, \sigma \bar{\sigma}}(x) c^\dagger_{a\bar{\sigma}k}(x) c_{a {\sigma} k}(x)\big] \\
        - U L_y &\displaystyle\sum_{x, a} \big[\chi_{aa, \uparrow\uparrow}(x) \chi_{aa, \downarrow\downarrow}(x) - |\chi_{aa, \uparrow \downarrow}(x)|^2 \big]&
    \end{aligned}
    \label{eq:MF_Hamiltonian}
\end{equation} 
with 
\begin{equation}
    \chi_{ab, \sigma \sigma'}(x) = \frac{1}{L_y} \displaystyle\sum_{k=1}^{L_y} \displaystyle\sum_{\lambda} \psi^*_{\lambda, a\sigma k}(x) \psi_{\lambda, b\sigma' k}(x) n_{k}^\lambda.
\end{equation}
We sum over all eigenstates $\lambda$ of the mean-field Hamiltonian, with eigenvalue $\varepsilon_{\lambda k}$ and eigenvector $\psi_{\lambda, a\sigma k}(x)$. Here, $n_k^\lambda = 1/(e^{\beta \varepsilon_{\lambda k}} + 1)$ is the Fermi-Dirac distribution at a temperature $T=\beta^{-1}$. We use the following workflow to obtain the domain-wall structures discussed in the main text: First, we use a uniform in $x$ initial state $\chi_{aa, \sigma \sigma}(x)$ and solve Eq.~\eqref{eq:MF_Hamiltonian} iteratively until both energy and $\chi$ are converged. From that, we obtain the system energy in the absence of domain walls $E_0$. We then use the converged uniform state to construct an initial state with two domain walls. Since we are working with periodic boundary conditions, the number of domain walls must be even. We then again solve Eq.~\eqref{eq:MF_Hamiltonian} iteratively, using the two-domain-wall initial state, see also Ref.~\cite{Qiu2025} for related calculations. This results in the domain-wall structure and surface tension presented in the main text. Numerically, we define the surface tension as
\begin{equation}
    \sigma = \frac{E-E_0}{2 L_y},
\end{equation}
where $E$ is the (free) energy with two domain walls, $E_0$ is the (free) energy without domain walls, and $L_y$ is the length of a single domain wall. 
Convergence is verified by introducing small random fluctuations to every initial state, which we find does not affect the converged result.
We treat the chemical potential as a Lagrange multiplier, which we adjust in each iteration to keep the particle number fixed. For the gapped quantum Hall states, we instead keep the chemical potential fixed (within the bulk gap).

\section{Tunable metric model and quantum metric} \label{app:TMmodel}
The tunable metric (TM) model~\cite{hofmannHeuristicBoundsSuperconductivity2022, hofmannTMmodel2023, Shavit_2025}
\begin{equation}
    H_0^\mathrm{TM} = H_t + \frac{E_g}{2} \displaystyle\sum_{\mathbf{k}} c^\dagger_\mathbf{k} [ \tau^x \sin(\zeta \alpha_\mathbf{k}) + \sigma^z \tau^y \cos(\zeta\alpha_\mathbf{k})]c_\mathbf{k}, \label{eq:TMmodel}
\end{equation}
 is defined on a square lattice and has an orbital degree of freedom, on which the Pauli matrices $\tau^i$ act, and a spin degree of freedom $\sigma^i$, resulting in a total of four bands. We have $\alpha_\mathbf{k} = \cos(k_x a) + \cos(k_ya)$, where $a$ is the lattice constant. In the limit $E_g \rightarrow \infty$, two of the four bands are effectively projected out, leaving us with two spin-degenerate bands. The dispersion of these bands is entirely determined by $H_t$, which we choose to be nearest-neighbor hoppings with amplitude $t$. The remaining terms in $H_\mathrm{TM}$ only affect the quantum geometry, which is freely tunable by the parameter $\zeta$. 

We introduce the density of states at the Fermi surface as
\begin{equation}
    N(0) = \int \frac{d^2q}{(2\pi)^2}\frac{\partial n_F(\xi_\mathbf{q})}{\partial\xi_\mathbf{q}}. \label{eq:DoS_at_FS}
\end{equation}
The Fermi-surface averaged quantum metric is given by
\begin{equation}
    \bar{g}_\mathrm{FS} = \int \frac{d^2q}{(2\pi)^2} \mathrm{Tr}[g_{ab}(\mathbf{q}) ] \frac{\partial n_F(\xi_\mathbf{q})}{\partial\xi_\mathbf{q}}. \label{eq:FermiAveragedMetric}
\end{equation}
The quantum metric $g_{ab}(\mathbf{q})$ has dimensions of $L^2$ (where $L$ is length), while in our convention the Fermi-surface averaged metric Eq.~\eqref{eq:FermiAveragedMetric} has dimensions of $E^{-1}$ (where $E$ is energy). 
For the TM model, where the quantum metric is given by~\cite{Shavit_2025}
\begin{equation}
    g_{ab}(\mathbf{q}) = \frac{(\zeta a)^2}{4} \sin(k_a a)\sin(k_b a),
\end{equation}
and the effective mass in the low density limit is $m^*=(2ta^2)^{-1}$, this results in
\begin{equation}
    \bar{g}_\mathrm{FS} = \frac{\zeta^2}{16\pi t} (k_F a)^2.
\end{equation}
Here we assumed that the Fermi momentum is small compared to the lattice spacing $k_F a \ll 1$.

\begin{figure}
\begin{center}
\includegraphics[width=0.99\linewidth]{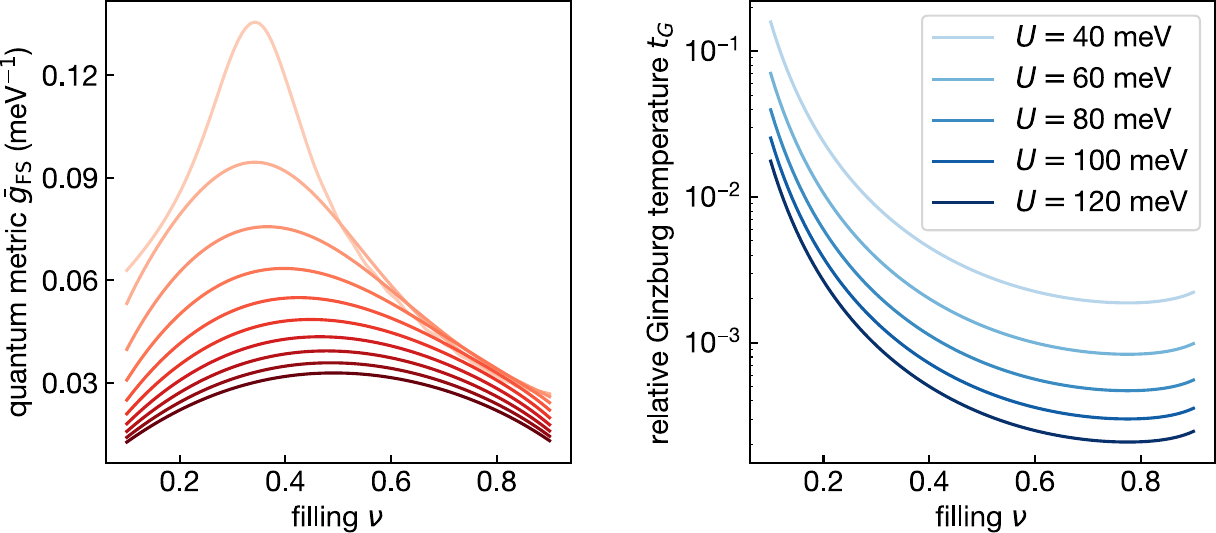}
\caption{\textbf{Ginzburg temperature and quantum metric in twisted MoTe$_2$.} Left: Fermi-surface averaged quantum metric for twisted MoTe$_2$ for different temperatures (from $T=1$ K for the lightest to $T=25$ K for the darkest shade). Right: Reduced Ginzburg temperature $t_G$ for twisted MoTe$_2$ at a twist angle of $\theta=3.7^\circ$ as a function of the filling $\nu$ of the topmost hole band, assuming different interaction strengths $U$.
}
\label{fig:GinzburgTemperatureMoTe2}
\end{center}
\end{figure}
\section{Model for twisted MoTe$_2$} \label{app:MoTe2}
To give an estimate for the reduced Ginzburg temperature in twisted MoTe$_2$, we use the following moir\'e Hamiltonian, which for the $K$ valley is given by~\cite{Wu_TopologicalInsulators_2019}
\begin{equation}
    {H}_0^\text{t-MoTe${}_2$} = \begin{pmatrix}
        - \frac{(\kk - \boldsymbol{\kappa}_+)^2}{2m^*}+\Delta_\bfr(\mathbf{r}) & \Delta_T(\mathbf{r}) \\
        \Delta^\dagger_T(\mathbf{r}) & - \frac{(\kk - \boldsymbol{\kappa}_-)^2}{2m^*}+\Delta_\tfr(\mathbf{r})
    \end{pmatrix} \label{eq:cont_2}
\end{equation}
with a layer $\ell \in \{\tfr=-1, \bfr=+1\}$ dependent moir\'e potential 
\begin{equation}
    \Delta_\ell(\mathbf{r}) =  \displaystyle\sum_{n \in \{0, 2, 4\}} 2 V\cos( \mathbf{G}_n \cdot \mathbf{r} + \ell \psi) \label{eq:moirepot_homo}
\end{equation}
and interlayer hopping term
\begin{equation}
    \Delta_T(\mathbf{r}) = w(1+e^{-i \mathbf{G}_1 \cdot \mathbf{r}}+e^{-i \mathbf{G}_2 \cdot \mathbf{r}}).
\end{equation}
Here $\mathbf{G}_n = \mathrm{R}_{n\pi/3} \mathbf{G}_0$ and $\mathbf{G}_0 = 4\pi/\sqrt{3} a_M (1, 0)$ are moir\'e reciprocal lattice vectors, with the moir\'e lattice constant $a_M = a/\theta$. We adapt the model parameters $(w, V, \psi, m^*)$ from Ref.~\cite{wang_FractionalChern_2024}. For the reduced Ginzburg temperature $t_G$, we need to compute the Fermi-surface averaged quantum metric Eq.~\eqref{eq:FermiAveragedMetric} (see Fig.~\ref{fig:GinzburgTemperatureMoTe2}). We numerically evaluate the quantum metric
\begin{equation}
        g_{ab}(\kk) = \frac{1}{2} \mathrm{Tr}\left( \frac{\partial P(\kk)}{\partial k_a} \frac{\partial P(\kk)}{\partial k_b} \right), \label{eq:quantummetric}
\end{equation}
for the topmost hole band with projector $P(\kk) = \mathcal{U}(\kk) \mathcal{U}^\dagger(\kk)$ and eigenvector $\mathcal{U}(\kk)$. The reduced Ginzburg temperature Eq.~\eqref{eq:relativeGinzburgTemp} depends on the effective interaction strength $U$, which in theoretical models of twisted MoTe$_2$ sensitively depends on the chosen dielectric screening constant~\cite{chen2025fractionalcherninsulatorquantum}. Hence, we calculate $t_G$ for several values of $U$ (see Fig.~\ref{fig:GinzburgTemperatureMoTe2}), assuming a critical temperature of approximately $T_c \approx 10$~K~\cite{anderson_ProgrammingCorrelated_2023, zeng_ThermodynamicEvidence_2023, park_ObservationFractionally_2023}. Due to the presence of a large quantum metric, which controls quantum fluctuations, we find generically small reduced Ginzburg temperatures $t_G \ll 1$, justifying our mean-field treatment in the main text.

To give order-of-magnitude estimates for the surface tension of twisted MoTe$_2$, it is convenient to work with a real-space lattice model, instead of the continuum model Eq.~\eqref{eq:cont_2}. The two topmost hole bands of Eq.~\eqref{eq:cont_2} can be mapped to an extended Haldane model on a honeycomb lattice with up to third-nearest neighbor hopping~\cite{chen2025fractionalcherninsulatorquantum, he2025fractionalcherninsulatorscompeting}. We use the following hopping parameters: $t_1=3.225$ meV, $t_2=-2.210$ meV, $t_3=-0.947$ meV. Nearest-neighbor and third-nearest hoppings are purely real, while the next-nearest neighbor hopping has a complex phase of $e^{i\frac{2\pi}{3}}$. These parameters are adopted from Ref.~\cite{chen2025fractionalcherninsulatorquantum} and approximate the particle-hole transformed two topmost bands of twisted MoTe$_2$ at a twist angle around $\theta=3.7^\circ$, corresponding to a moir\'e lattice constant of $a_M \approx 5.5$ nm. Upon increasing repulsive interactions, the model transitions to a ferromagnetic phase which breaks discrete time reversal symmetry $U(1)_{S_z}\times \mathcal{T} \rightarrow U(1)_{S_z}$. The presence of topological bands leads to an Ising anisotropy for the magnet~\cite{Wu_TopologicalInsulators_2019,Qiu2025}, which allows us to neglect spin-wave fluctuations in our model. For clarity, we restrict the model in the main text to repulsive on-site interactions $U$. In a more accurate description of twisted MoTe$_2$, longer-range interactions are expected to be relevant. However, we have verified that including interaction terms beyond the on-site term does not qualitatively modify our results. Since our goal is to obtain an order-of-magnitude estimate, and given that it is not clear to what extent a honeycomb-lattice model is sufficient to quantitatively describe twisted MoTe$_2$, we choose to work with the minimal model for simplicity. 
We assume translational invariance in the $y$ direction, while we put two domain walls in the $x$ direction. 
In the zero-temperature limit for $U=50\; \mathrm{meV}$, we obtain a surface tension of $\sigma_0 \approx 0.17 \; \mathrm{meV/nm}$, corresponding to a critical bubble radius of $R_c \approx 2 \; \mu\mathrm{m}$ under a magnetic field of $B_z = 10\; \mathrm{mT}$ (to be compared with typical coercive fields of $B_z^c \sim  100\; \mathrm{mT}$). While varying $U$ or including longer-range interactions will quantitatively alter the value of the surface tension $\sigma$, the key takeaway is that the expected critical radius is comparable to an optical spot size.

\clearpage

\onecolumngrid

\setcounter{equation}{0}
\setcounter{figure}{0}
\setcounter{table}{0}
\setcounter{page}{1}

\renewcommand{\theequation}{S\arabic{equation}}
\renewcommand{\thefigure}{S\arabic{figure}}
\renewcommand{\thetable}{S\arabic{table}}

\begin{center}
    \textbf{\large Supplemental Material: \\ False Vacuum Decay in Flat-Band Ferromagnets: Role of Quantum Geometry and Chiral Edge States}
\end{center}

\begin{center}
    Fabian Pichler, Clemens Kuhlenkamp, and Michael Knap
\end{center}

\section{Static particle-hole bubble with quantum geometry}

Here, we show that the static particle-hole bubble
\begin{equation}
    \Pi(\mathbf{q}) = -i \displaystyle\int_0^\infty dt \langle [ \rho_{\mathbf{q}}(t), \; \rho_{-\mathbf{q}}(0)] \rangle,
\end{equation}
which for a quadratically dispersing band in two dimensions is constant without quantum metric for $|\mathbf{q}|<2k_F$~\cite{KITTEL19691},
picks up a zero-momentum curvature in bands with non-zero quantum metric (see Eq.~\ref{eq:staticPolarization}). We project to the lowest band of $H_0$ in Eq.~\eqref{eq:generalHamiltonian}, where the electron operator is given by
\begin{equation}
    \gamma^\dagger_\mathbf{k} = \displaystyle\sum_{a} \mathcal{U}_{a}(\mathbf{k}) c^\dagger_{a\mathbf{k}}.
\end{equation}
For notational simplicity, we are suppressing spin indices. The projected density operator is
\begin{equation}
    \rho_\mathbf{q} = \frac{1}{\sqrt{N}} \displaystyle\sum_{\mathbf{k}} \Lambda_{\mathbf{k}+\mathbf{q}, \mathbf{k}} \gamma^\dagger_{\mathbf{k}+\mathbf{q}} \gamma_{\mathbf{k}}
\end{equation}
 with form factors $\Lambda_{\mathbf{k}, \mathbf{k}'} = \sum_a \mathcal{U}^*_{a}(\mathbf{k}) \mathcal{U}_{a}(\mathbf{k}')$. The free static particle-hole bubble then becomes
 \begin{equation}
     \Pi(\mathbf{q}) = \frac{1}{N} \displaystyle\sum_\mathbf{k} \frac{n_F(\xi_{\mathbf{k}+\mathbf{q}}) - n_F(\xi_{\mathbf{k}})}{\xi_{\mathbf{k}+\mathbf{q}}- \xi_{\mathbf{k}}} |\Lambda_{\mathbf{k}+\mathbf{q}, \mathbf{k}}|^2.
    \label{eq:PibubbleAppendix}
 \end{equation}
 where $\xi_\kk \approx |\mathbf{k}|^2/2m^* - \mu$ is assumed to be approximately quadratic and $n_F$ is the Fermi-Dirac distribution. 
 Next, we expand $|\Lambda_{\mathbf{k}+\mathbf{q}, \mathbf{k}}|^2$ to quadratic order in $q$. To do that, it is convenient to write 
 \begin{equation}
     |\Lambda_{\mathbf{k}+\mathbf{q}, \mathbf{k}}|^2 = \mathrm{Tr}[P(\mathbf{k}+\mathbf{q})P(\mathbf{k})],
 \end{equation}
 with projector $P(\kk) = \mathcal{U}(\kk) \mathcal{U}^\dagger(\kk)$. Using the projector properties $P^2(\kk) = P(\kk)$ and $\mathrm{Tr}[P(\kk)]=1$, we find that 
 \begin{equation}
     \begin{aligned}
         \mathrm{Tr}[P(\kk)\partial_a P(\kk)] &= 0 \quad \mathrm{and} \\
         \mathrm{Tr}[P(\kk)\partial_a \partial_b P(\kk)] &= -\mathrm{Tr}[\partial_aP(\kk) \partial_b P(\kk)],
     \end{aligned}
 \end{equation}
with $\partial_a = \partial/\partial k_a$. From these identities, it follows that
 \begin{equation}
     |\Lambda_{\mathbf{k}+\mathbf{q}, \mathbf{k}}|^2 = 1 - \displaystyle\sum_{a b}g_{ab}(\kk) q_a q_b + \mathcal{O}(q^3)
 \end{equation}
 with the quantum metric $g_{ab}$ defined in Eq.~\eqref{eq:quantummetric}. Replacing the sum in Eq.~\eqref{eq:PibubbleAppendix} by an integral and taking the $q\rightarrow0$ limit, we find
 \begin{equation}
     \Pi(\mathbf{q}) \approx \int \frac{d^2k}{(2\pi)^2}  \frac{\partial n_F(\xi_\mathbf{k})}{\partial\xi_\mathbf{k}} \Big[ 1 - \displaystyle\sum_{a b}g_{ab}(\kk) q_a q_b \Big]. \
 \end{equation}
 Using the definition Eq.~\eqref{eq:FermiAveragedMetric} of the Fermi-surface averaged quantum metric, we obtain Eq.~\eqref{eq:staticPolarization} from the main text. Note that we have assumed that the static particle-hole bubble has no zero-momentum curvature without quantum metric, which holds for bands in two dimensions with quadratic dispersion~\cite{KITTEL19691}.

\section{Cross-check of Landau-Ginzburg Approach}

\begin{figure}
\begin{center}
    \includegraphics[width=0.6\linewidth]{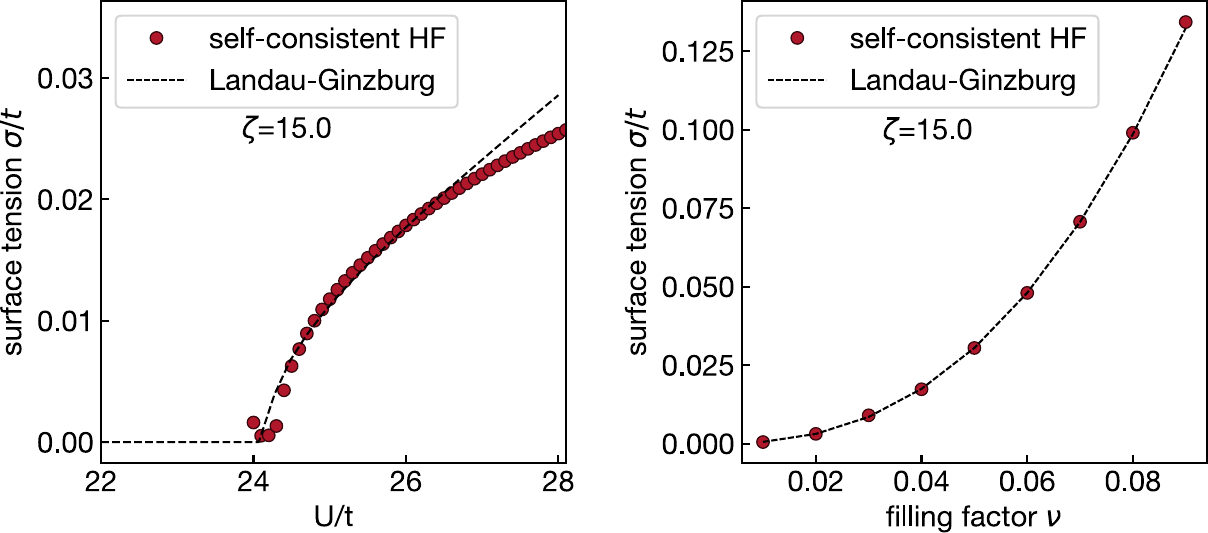}
    \caption{\textbf{Comparison of surface tension from Landau-Ginzburg theory and self-consistent Hartree-Fock}. Left: Surface tension as a function of interaction strength $U$, across the Stoner transition in the TM model. Right: Dependence of surface tension on the electron density (expressed through the filling factor $\nu$), for fixed $U=25t$.}
    \label{fig:cross-check}
\end{center}
\end{figure}

As a cross-check of our Landau-Ginzburg treatment in main text, we compare it with the surface tension determined from self-consistent Hartree-Fock.
The Landau-Ginzburg surface tension of the TM-model, Eq.~\eqref{eq:TMmodel}, follows from follows from Eq.~\eqref{eq:surfacetension} and is given by
\begin{equation}
    \sigma = \frac{4}{3} m_0^2 \sqrt{2U^3 \big(U N(0) -1\big) \bar{g}_\mathrm{FS} }. \label{eq:surfaceTMmodel}
\end{equation}
We find good quantitative agreement between 
Eq.~\eqref{eq:surfaceTMmodel} and self-consistent Hartree-Fock; see Fig.~\ref{fig:cross-check}.

\end{document}